\documentclass[runningheads]{llncs}
\usepackage{type1cm}        % activate if the above 3 fonts are
\usepackage{makeidx}         % allows index generation
\usepackage{graphicx}        % standard LaTeX graphics tool
\usepackage{subcaption}                             % when including figure files
\usepackage{multicol}        % used for the two-column index
\usepackage[bottom]{footmisc}% places footnotes at page bottom

\usepackage{orcidlink}
\usepackage{newtxtext}       % 
\usepackage{newtxmath}       % selects Times Roman as basic font

\usepackage{amsmath}
\usepackage{algorithm}
\usepackage{algpseudocode}
\usepackage{graphicx}

\usepackage[version=4]{mhchem}
\usepackage{siunitx}
\usepackage{longtable,tabularx}
\usepackage{subfiles} % Best loaded last in the preamble
\usepackage{pdfpages} %Inserts pdf pages
\usepackage{csvsimple} % Imports csv documents as pdf pages
\usepackage{tikz} %prepares flowcharts
\usepackage{import}            % for \subimport
\usetikzlibrary{quotes,arrows.meta}
\usetikzlibrary{positioning}

\usepackage{Ball}
\usepackage{Box}
\usepackage{RightBandedBox}

\usetikzlibrary{shapes, arrows, positioning} %imports shapes form library.
\usepackage{xurl} 
\usepackage{hyperref}

\usepackage[T1]{fontenc}
\begin{document}
\title{Rule-Based Spatial Mixture-of-Experts U-Net for Explainable Edge Detection}

\titlerunning{Explainable Rule-Based sMoE Edge Detection}
\author{Bharadwaj Dogga\inst{1}\orcidlink{0009-0002-5915-1367} \and
Kaaustaaub Shankar\inst{2}\orcidlink{0009-0002-5915-1367} \and
Gibin Raju\inst{3}\orcidlink{0000-0003-2559-6931} \and
Wilhelm Louw\inst{4}\orcidlink{0009-0009-8683-1932} \and
Kelly Cohen\inst{5}\orcidlink{0000-0002-8655-1465}}
\authorrunning{B. Dogga et al.}
\institute{Ph.D. Candidate, AI Bio Lab, Digital Futures, University of Cincinnati, \\Cincinnati, OH 45221, USA e-mail:\email{doggabj@mail.uc.edu} \and
Research Student, AI Bio Lab, Digital Futures, University of Cincinnati, \\Cincinnati, OH 45221, USA e-mail:\email{shankaks@mail.uc.edu} \and
Postdoctoral Researcher, Department of Multidisciplinary Engineering, TA\&M University, \\College Station, TX 77843, USA e-mail: \email{gibinraju@tamu.edu} \and
Research Associate AI Bio Lab, Digital Futures, University of Cincinnati, \\Cincinnati, OH 45221, USA e-mail:\email{louwwa@ucmail.uc.edu} \and
Director AI Bio Lab, Digital Futures, University of Cincinnati, \\Cincinnati, OH 45221, USA e-mail:\email{cohenky@ucmail.uc.edu}
}
\maketitle              % typeset the header of the contribution
\begin{abstract}

Deep learning models like U-Net and its variants, have established state-of-the-art performance in edge detection tasks and are used by Generative AI services world-wide for their image generation models. However, their decision-making processes remain opaque, operating as "black boxes" that obscure the rationale behind specific boundary predictions. This lack of transparency is a critical barrier in safety-critical applications where verification is mandatory. To bridge the gap between high-performance deep learning and interpretable logic, we propose the Rule-Based Spatial Mixture-of-Experts U-Net (sMoE U-Net). Our architecture introduces two key innovations: (1) Spatially-Adaptive Mixture-of-Experts (sMoE) blocks integrated into the decoder skip connections, which dynamically gate between "Context" (smooth) and "Boundary" (sharp) experts based on local feature statistics; and (2) a Takagi-Sugeno-Kang (TSK) Fuzzy Head that replaces the standard classification layer. This fuzzy head fuses deep semantic features with heuristic edge signals using explicit IF-THEN rules. We evaluate our method on the BSDS500 benchmark, achieving an Optimal Dataset Scale (ODS) F-score of 0.7628, effectively matching purely deep baselines like HED (0.7688) while outperforming the standard U-Net (0.7437). Crucially, our model provides pixel-level explainability through "Rule Firing Maps" and "Strategy Maps," allowing users to visualize whether an edge was detected due to strong gradients, high semantic confidence, or specific logical rule combinations.

\end{abstract}

\section{Introduction}
\label{sec:1}

Edge detection is a fundamental challenge in computer vision, serving as the precursor to image segmentation, object recognition, and scene understanding. The evolution of this field has been marked by a trade-off between explainability and accuracy\cite{teng2024literature}. Classical spatial methods, such as the Canny and Sobel operators, offered complete transparency as engineers could mathematically define exactly why a pixel was classified as an edge based on gradient thresholds. However, these systems lacked the semantic understanding to distinguish between true object boundaries and texture noise\cite{lynn2021implementation}.

Conversely, modern Convolutional Neural Networks (CNNs), such as the U-Net\cite{ronneberger_u_net_2015} and Holistically-Nested Edge Detection (HED)\cite{xie2015holistically}, have achieved remarkable accuracy by learning hierarchical feature representations. Yet, this performance comes at the cost of opacity. When a CNN fails, for instance, by hallucinating an edge in a smooth region, diagnosing the failure is non-trivial due to the distributed nature of the learned weights\cite{cosgrove2020adversarial}. To address these shortcomings, this paper proposes a hybrid neuro-fuzzy architecture. We introduce the explainable sMoE U-Net, a system that retains the semantic feature extraction power of Deep Learning while transparency through fuzzy logic. Our contribution includes:

\begin{enumerate}
    \item Spatially-Adaptive MoE: We implement a Mixture of Expert switch  that dynamically assigns pixels to either a "Smooth Expert" (for context) or a "Sharp Expert" (for localization), explicitly modeling the trade-off between semantic suppression and boundary preservation.
    \item First-Order TSK Fuzzy Head: Unlike standard neural networks that end in a sigmoid activation, our network terminates in a differentiable TSK fuzzy inference engine. This layer makes decisions based on learned rules (e.g., "IF Edge Strength is High AND Semantic Confidence is High THEN Edge is Present").
    \item Explainable Visualizations: We demonstrate that our model produces human-readable Strategy Maps and Rule Firing Maps, providing a granular view of the decision logic.
\end{enumerate}

\subsection{Related Work}
\subsubsection{Classical and Deep Edge Detection:}
Early work in edge detection focused on identifying discontinuities in image intensity. The Canny edge detector, remains a standard baseline due to its use of hysteresis thresholding to enforce edge continuity. While effective for simple scenes, Canny struggles with texture and requires manual tuning of thresholds.

The deep learning era introduced fully convolutional networks (FCNs) that view edge detection as a pixel-wise binary classification task\cite{su_pixel_nodate}. The U-Net architecture, originally designed for biomedical segmentation, utilizes skip connections to preserve spatial information lost during downsampling. More specialized architectures like HED utilize multi-scale feature fusion to capture both fine and coarse edges simultaneously.
\subsubsection{Mixture of Experts:} Mixture of Experts (MoE)\cite{dryden2022spatial} decomposes complex learning tasks into sub-tasks handled by specialized expert networks, mediated by a gating mechanism. While recently popularized in Natural Language Processing (NLP) via sparse-gating to scale model capacity\cite{cai2025survey}, MoE in computer vision has historically been utilized to handle scale variance or multi-task learning\cite{riquelme2021scaling}. 
\subsubsection{Neuro-Fuzzy Systems in Vision:}
Fuzzy logic systems excel at modeling uncertainty and vagueness, making them ideal for handling the ambiguity of "soft" edges\cite{rahim_towards_2025}. Historically, Adaptive Neuro-Fuzzy Inference Systems (ANFIS) were used for simple thresholding tasks. However, these systems struggled to scale to high-dimensional image data. Recently, there has been a resurgence in Deep fuzzy systems that integrate fuzzy layers into CNN backbones\cite{boskovitz_adaptive_2002}. Our work advances this trend by replacing the static aggregation methods of previous works with a dynamic Mixture-of-Experts approach.

\section{Methodology}
\label{sec:2}

\begin{figure}[htbp]
    \centering
    \includegraphics[width=\textwidth, page=1]{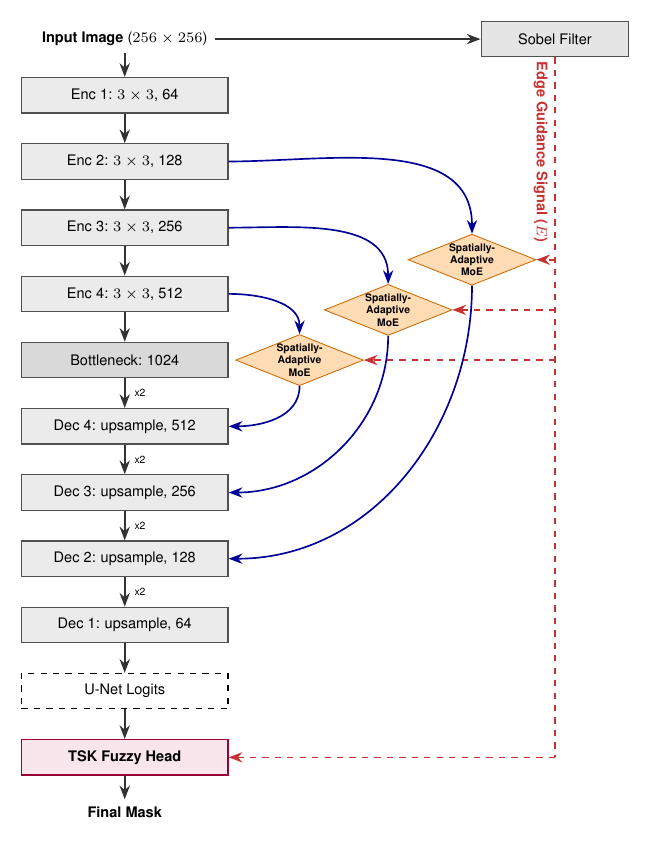}
    \caption{Compact architecture of the proposed explainable sMoE U-Net with Sobel pre-processing and a TSK fuzzy head.}
    \label{fig:explainable_UNet_arch}
\end{figure}

Our proposed architecture, the sMoE U-Net, modifies the standard Encoder-Decoder U-Net by injecting logic-driven components at two critical stages: feature refinement (via sMoE) and final classification (via TSK Fuzzy Head). This hybrid design is illustrated in Figure \ref{fig:explainable_UNet_arch}, which shows the placement of sMoE blocks at the decoder skip connections and the substitution of the classic sigmoid head with our fuzzy inference engine.

\subsection{Spatially-Adaptive Mixture-of-Experts (sMoE)}
\begin{figure}[htbp]
    \centering
    \includegraphics[width=\textwidth, page=2]{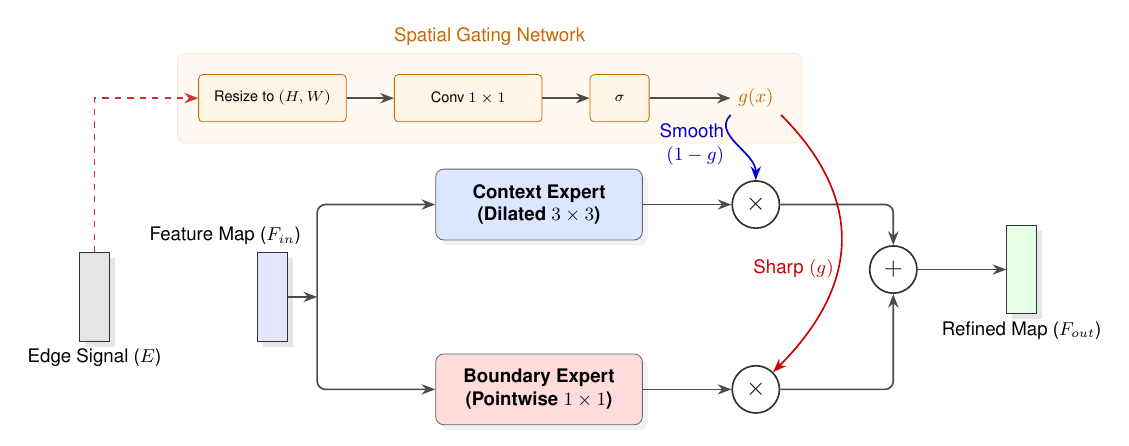}
    \caption{Architecture of Spatially-Adaptive Mixture-of-Experts with Sobel Edge signal}
    \label{fig:sMoE_arch}
\end{figure}

Standard U-Net skip connections simply concatenate encoder features with decoder features. This treats all pixels equally, regardless of whether they belong to a homogeneous region or a sharp boundary, often leading to the propagation of noise from early encoder layers. To address this, we introduce Spatially-Adaptive Mixture-of-Experts (sMoE), detailed in Figure \ref{fig:sMoE_arch}.

For a given input feature map $X$ and a heuristic edge map $E$ (derived from Sobel filtering), the sMoE computes the output $Y$ as a weighted sum of two experts:

\begin{equation}
    Y = G(E) \cdot \text{Expert}_{sharp}(X) + (1 - G(E)) \cdot \text{Expert}_{smooth}(X)
    \label{equation_1}
\end{equation}

where the gating signal $G(E)$ determines the pixel-wise processing strategy.

\begin{itemize}
    \item The Context Expert (Smooth): This branch utilizes dilated $3\times3$ convolutions (dilation=2) to expand the receptive field . By "looking around" the central pixel, this expert captures regional context, effectively suppressing texture noise in homogeneous areas.
    \item The Boundary Expert (Sharp): This branch utilizes pointwise $1\times1$ convolutions. This operation focuses exclusively on the local pixel intensity, preserving high-frequency spatial details and ensuring that edge boundaries remain crisp and unblurred.
    \item The Gating Network: As shown in Figure \ref{fig:sMoE_arch}, the gating network consists of a learnable $1\times1$ convolution that processes the resized edge guidance signal $E$. It produces a soft mask $G(E) \in [0, 1]$, which allows the network to differentiably select between smoothing (context) and sharpening (boundary) strategies for every individual pixel.
\end{itemize}

\subsection{TSK Fuzzy Explainability Head}
\begin{figure}[htbp]
    \centering
    \includegraphics[width=\textwidth, page=3]{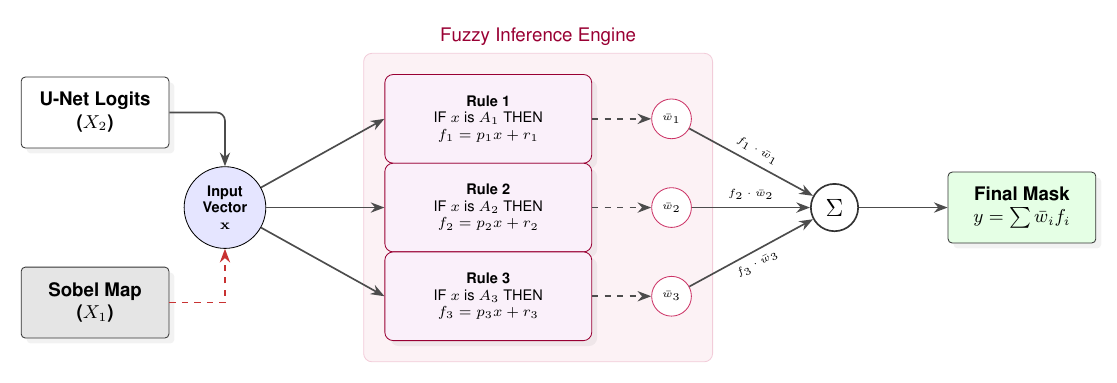}
    \caption{Architecture of proposed TSK Rule-Based head}
    \label{fig:fuzzy_unet}
\end{figure}

To render the final decision transparent, we replace the standard convolutional head with a Takagi-Sugeno-Kang (TSK) Fuzzy Inference Engine. As depicted in Figure \ref{fig:fuzzy_unet}, this head accepts a hybrid input vector consisting of:
\begin{enumerate}
    \item $x_1$ (Edge Strength): The normalized magnitude from a fixed Sobel layer.
    \item $x_2$ (Semantic Confidence): The sigmoid activation of the U-Net's deep features.
\end{enumerate}
The system learns $R=4$ rules, enabling it to model complex, non-linear edge definitions. The $i$-th rule is defined as a First-Order TSK rule:

$$\textbf{Rule } i: \text{IF } x_1 \text{ is } A_{i,1} \text{ AND } x_2 \text{ is } A_{i,2} \text{ THEN } y_i = a_{i,0} + a_{i,1}x_1 + a_{i,2}x_2$$

Where $A_{i,j}$ are Gaussian membership functions defined by trainable centers $c_{i,j}$ and widths $\sigma_{i,j}$. The firing strength $w_i$ represents the degree to which a rule applies to the current pixel, calculated as the product of the memberships: $w_i = \mu_{i,1}(x_1) \cdot \mu_{i,2}(x_2)$. The final predicted edge map $Y_{final}$ is computed as the weighted average of the rule consequents, ensuring the output is fully differentiable and interpretable.

\section{Experimental Setup}
\label{sec:3}

\subsection{Dataset and Implementation}

We utilize the BSDS500 dataset\cite{MartinFTM01}, consisting of 200 training, 100 validation, and 200 test images. The training set is augmented via rotation to improve robustness of the UNet backbone\cite{ocak_iocak28unet_edge_detection_2024}. The model is implemented in PyTorch and trained on a single NVIDIA L4 GPU on Google Colab. We used Adam optimizer with a learning rate of $1e^{-4}$ and a batch size of 4.
\subsection{Loss Function}
The model is trained using a composite loss function shown in equation \ref{eq:loss_equation} that combines Binary Cross Entropy (BCE) and Dice Loss to handle class imbalance. In this case, the imbalance is denoted by sparsity of edges.
\begin{equation}
    \mathcal{L}_{total} = \lambda \mathcal{L}_{BCE} + (1-\lambda) \mathcal{L}_{Dice}
    \label{eq:loss_equation}
\end{equation}
For the explainable head, we employ a distillation approach where the fuzzy output is trained to mimic the main U-Net logits using MSE loss.

\section{Results}
\label{sec:4}

The network demonstrated stable convergence behavior on the BSDS500 dataset. The final training loss stabilized at $0.3519$, while the validation loss settled at $0.4254$. This narrow generalization gap suggests that the model effectively learned robust feature representations without suffering from severe overfitting. The successful minimization of the total loss indicates that the composite objective function enabled the sMoE U-Net to capture both fine structural details and global edge continuity.

% Figure \ref{fig:sMoE_UNET_results} shows the edge detection results. 

% \begin{figure}[htbp]
%     \centering
%     \includegraphics[trim=0cm 0.0cm 0cm 0.2cm, clip, width=\linewidth]{images/results_images/exp_u-net_results.png}
%     \caption{Edge detection results from sMoE explainable UNeT}
%     \label{fig:sMoE_UNET_results}
% \end{figure}

\subsection{Quantitative Comparison}

We evaluate the sMoE U-Net on the BSDS500 benchmark, comparing it against classical operators (Canny, Sobel) and deep learning baselines (U-Net, HED). Table \ref{tab:quantative_metrics} summarizes the performance on test set.
\begin{table}[h!]
\centering
\begin{tabular}{l|c|c|c}
%\hline
\textbf{Method} & \textbf{OIS} & \textbf{ODS} & \textbf{AP} \\
\hline
&&\\
U-Net     & 0.7260 & 0.7437 & 0.6946  \\
sMoE U-Net (Ours) & 0.7458 & 0.7628 & \textbf{0.7222} \\
Canny     & 0.4836 & 0.5450 & 0.4125 \\
Sobel       & 0.5303 & 0.5769 & 0.4743 \\
HED  & \textbf{0.7514} & \textbf{0.7688} & 0.7126
\end{tabular}
\caption{Quantitative Evaluation of Edge Detection Methods}
\label{tab:quantative_metrics}
\end{table}

The classical methods struggle significantly. Canny achieves an ODS F-score of only 0.5450, hampered by its inability to distinguish texture from structure. The standard U-Net baseline demonstrates the value of learned features, reaching an ODS of 0.7437. However, our explainable sMoE U-Net outperforms the standard U-Net across all metrics, achieving an ODS of 0.7628 and an OIS of 0.7458. This performance gain validates the efficacy of the sMoE blocks in refining feature representations. Crucially, our model achieves an Average Precision (AP) of 0.7222, surpassing both the U-Net (0.6946) and the specialized HED architecture (0.7126), indicating superior stability across varying recall thresholds.

\begin{figure}[htbp]
    \centering
    \includegraphics[trim=0cm 0.3cm 0cm 0.7cm, clip, width=0.6\linewidth]{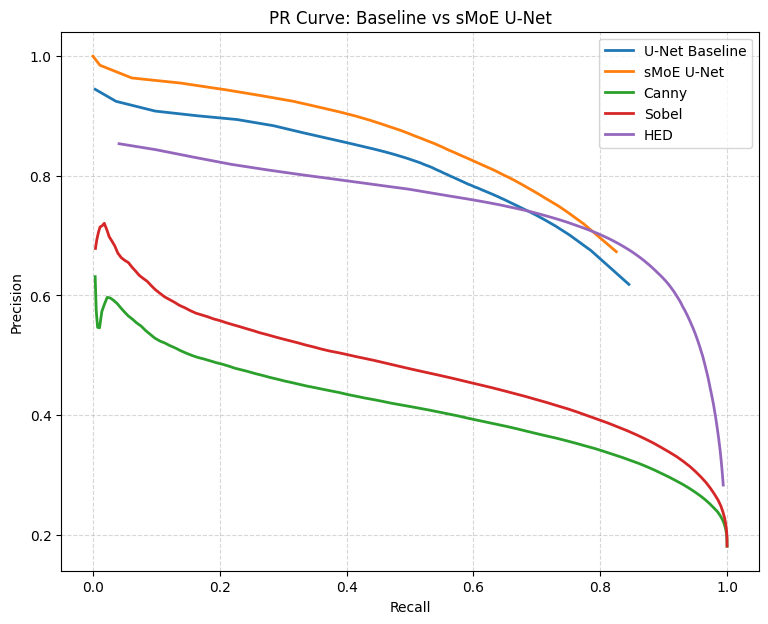}
    \caption{Precision-Recall Curve}
    \label{fig:pr_curve}
\end{figure}

These findings are further corroborated by the Precision-Recall curve in Figure \ref{fig:pr_curve}. The sMoE U-Net (orange curve) consistently maintains higher precision at equivalent recall levels compared to the U-Net baseline (blue curve), demonstrating that our sMoE-driven gating effectively reduces false positives without sacrificing edge discovery.

\subsection{Qualitative and explainability Analysis}
\begin{figure}[htbp]
    \centering
    
    % First Subfigure
    \begin{subfigure}[b]{\linewidth}
        \centering
        \includegraphics[width=\linewidth]{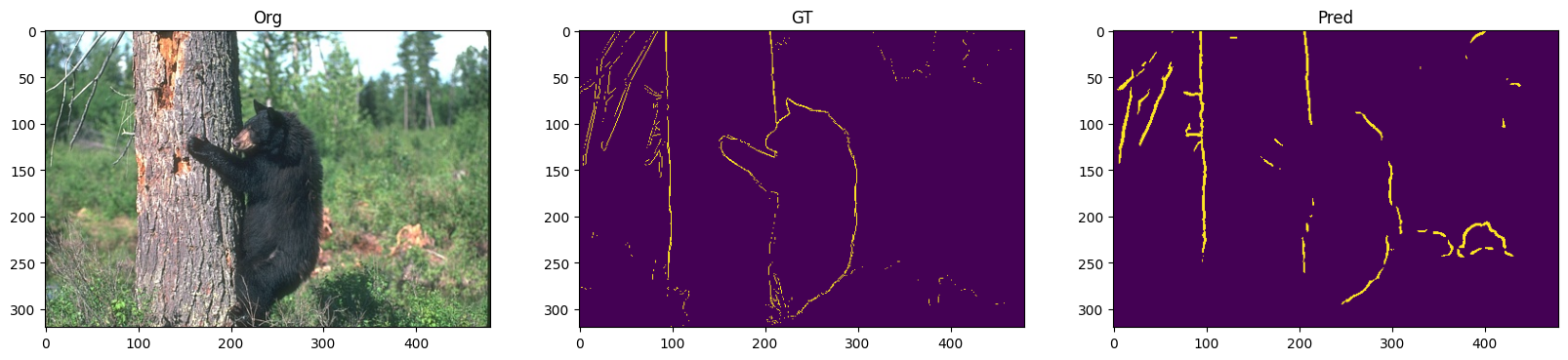}
        \caption{U-Net}
        \label{fig:unet_results}
    \end{subfigure}
    
    \vspace{0.5cm} % Add vertical space between figures

    % Second Subfigure
    \begin{subfigure}[b]{\linewidth}
        \centering
        \includegraphics[width=\linewidth]{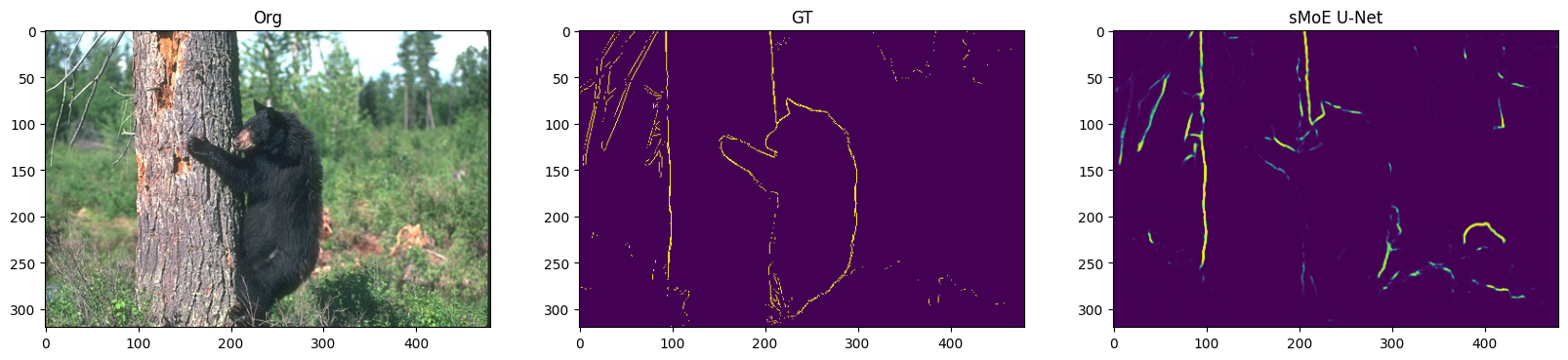}
        \caption{sMoE U-Net}
        \label{fig:explainable_sMoE_UNet_results}
    \end{subfigure}

    \vspace{0.5cm}

    % Third Subfigure
    \begin{subfigure}[b]{\linewidth}
        \centering
        \includegraphics[width=\linewidth]{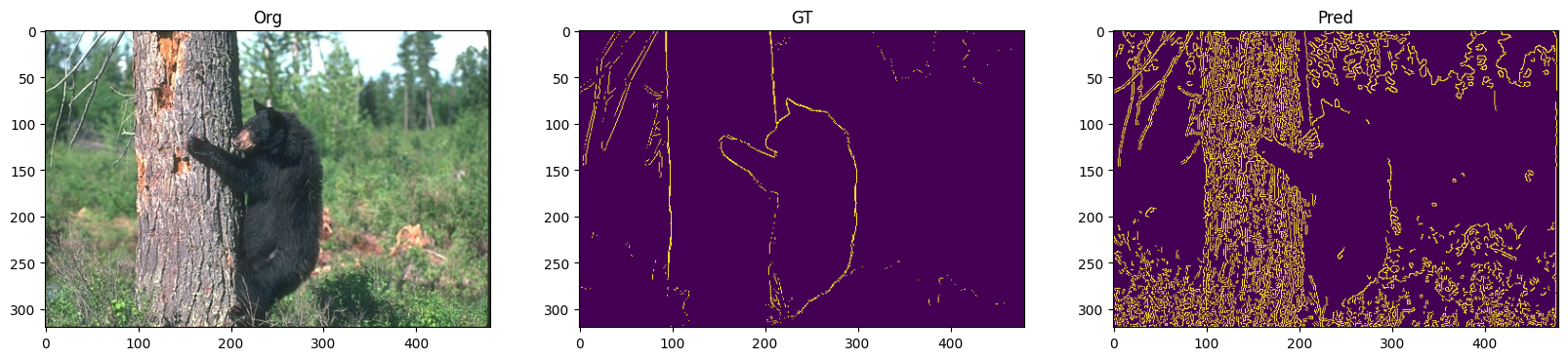}
        \caption{Canny Filter}
        \label{fig:canny_results}
    \end{subfigure}

    \vspace{0.5cm}

    % Fourth Subfigure
    \begin{subfigure}[b]{\linewidth}
        \centering
        \includegraphics[width=\linewidth]{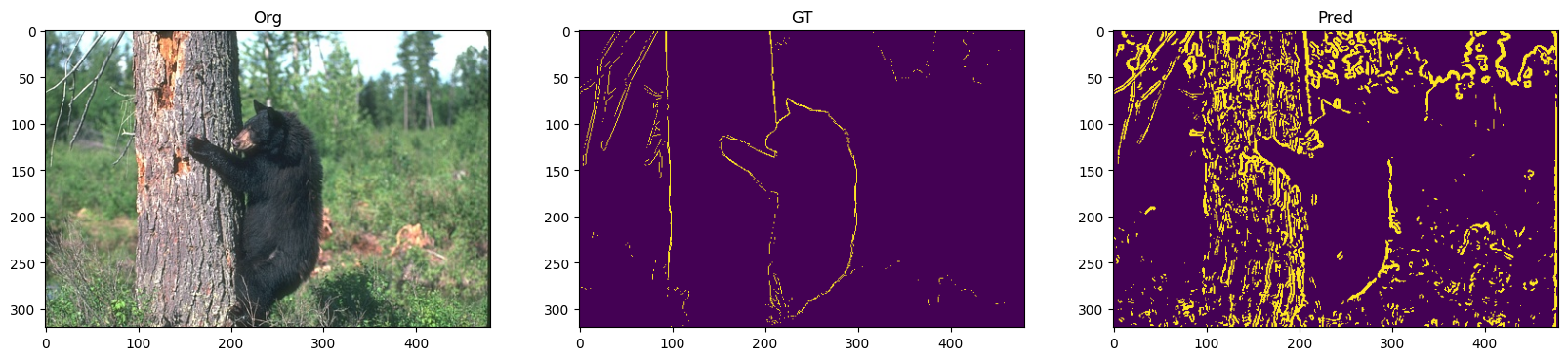}
        \caption{Sobel Filter}
        \label{fig:sobel_results}
    \end{subfigure}

    \vspace{0.5cm}

    % Fifth Subfigure
    \begin{subfigure}[b]{\linewidth}
        \centering
        \includegraphics[width=\linewidth]{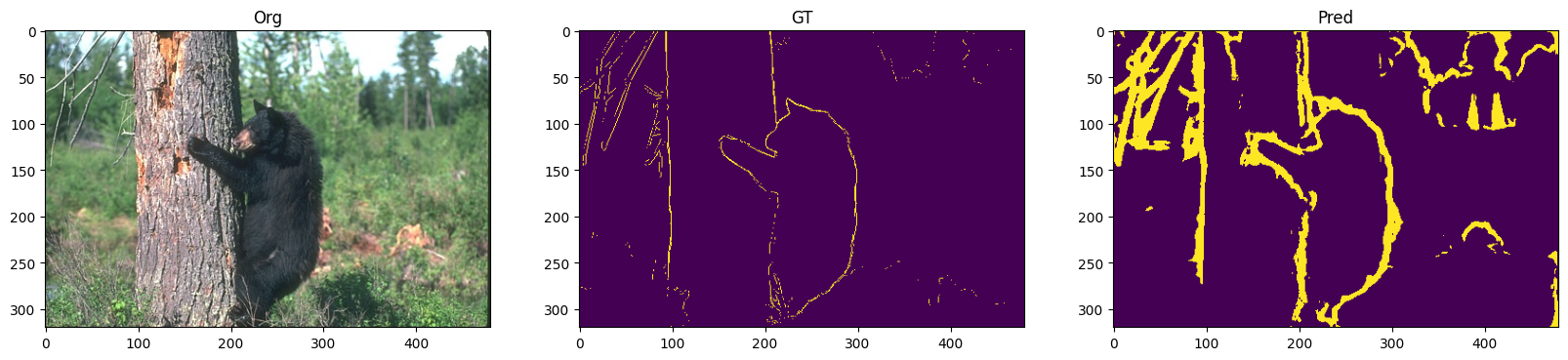}
        \caption{Holistically-Nested Edge Detection}
        \label{fig:hed_results}
    \end{subfigure}

    \caption{Edge detection comparison across various methods}
    \label{fig:comparison_results}
\end{figure}

Figure \ref{fig:comparison_results} provides a comprehensive visual comparison. While the Sobel filter \ref{fig:sobel_results} produces thick, noisy edges and Canny \ref{fig:canny_results} fails to capture the semantic outline of the bear, often hallucinating edges in the grassy texture, the sMoE U-Net \ref{fig:explainable_sMoE_UNet_results} generates a crisp, clean boundary map that closely aligns with the Ground Truth (GT).

% \begin{figure}[htbp]
%     \centering
%     % Replace 'HED_pred_results.jpg' with the exact filename you uploaded.
%     % [width=\linewidth] scales the image to exactly the width of the text block.
%     \includegraphics[trim=0cm 0.3cm 0cm 0.7cm, clip, width=0.7\linewidth]{images/results_images/training_val_loss.png}
    
%     \caption{Add your descriptive caption here.}
%     \label{fig:hed_results}
% \end{figure}

\begin{figure}[htbp]
    \centering

    \includegraphics[trim=0cm 0.0cm 0cm 0.0cm, clip, width=\linewidth]{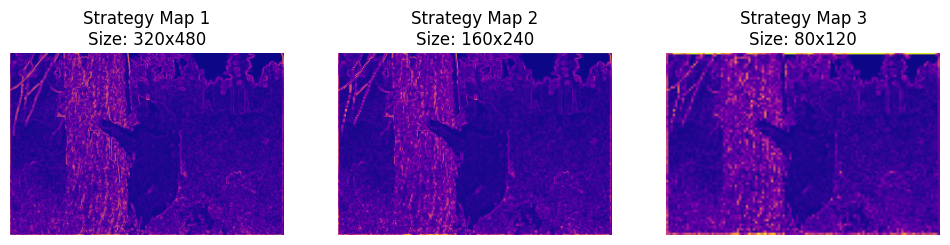}
    
    \caption{Strategy Maps across layers}
    \label{fig:strategy_maps}
\end{figure}

\begin{figure}[htbp]
    \centering

    \includegraphics[trim=0cm 0cm 0cm 0cm, clip, width=\linewidth]{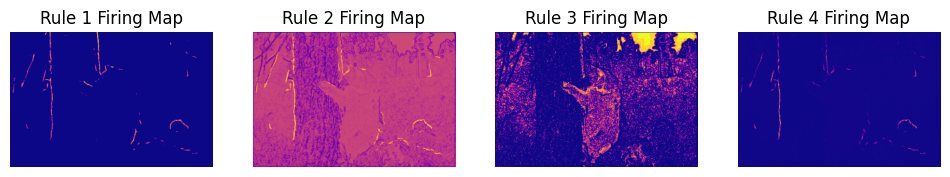}
    
    \caption{Fuzzy Rule-Base firing maps}
    \label{fig:tsk_rule_firing_map}
\end{figure}

Beyond accuracy, our model offers unique visual explanations. Figure \ref{fig:strategy_maps} displays the "Strategy Maps" generated by the gating network. These maps reveal that the model actively assigns different processing strategies to different image regions: the "Boundary Expert" (bright regions) is activated specifically along the bear's outline, while the "Context Expert" (dark regions) is utilized for the background, validating our hypothesis of spatially-adaptive processing. Furthermore, Figure \ref{fig:tsk_rule_firing_map} visualizes the "Rule Firing Maps" from the fuzzy head. Distinct rules can be seen firing for different edge types. For instance, one rule activates on strong, unambiguous edges, while another captures fainter semantic boundaries, thus providing engineers with a granular view of the model's decision logic.

\begin{figure}[htbp]
    \centering
    \begin{subfigure}[b]{0.45\textwidth}
        \centering
        % Replace 'example-image-a' with your first image filename
        \includegraphics[trim=0cm 0cm 0cm 0.8cm, clip, width=\textwidth]{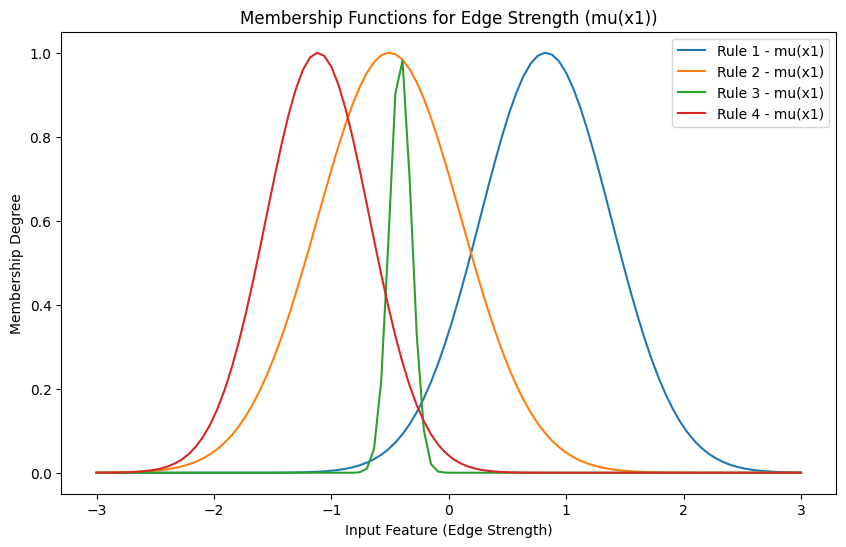}
        \caption{Trained rules for Spatial Confidence}
        \label{fig:x1_visual}
    \end{subfigure}
    \hfill % Adds flexible space between the two top images
    % Subfigure 2 (Top Right)
    \begin{subfigure}[b]{0.45\textwidth}
        \centering
        % Replace 'example-image-b' with your second image filename
        \includegraphics[trim=0cm 0cm 0cm 0.8cm, clip, width=\textwidth]{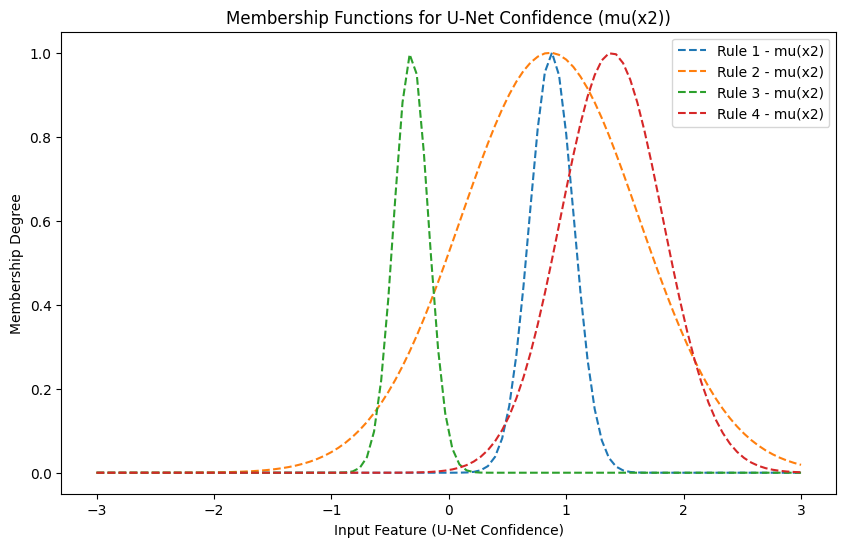}
        \caption{Trained rules for Semantic Confidence}
        \label{fig:x2_visual}
    \end{subfigure}

    % --- Forced Line Break and Vertical Space ---
    % \par ends the paragraph to force a new line.
    % \bigskip adds some vertical space between rows.
    \par\bigskip

    % --- Bottom Row: One image centered below ---
    % Subfigure 3 (Bottom)
    % You can adjust the width here (e.g., {0.8\textwidth} for a wider bottom image)
    \begin{subfigure}[b]{0.6\textwidth}
        \centering
        % Replace 'example-image-c' with your third image filename
        \includegraphics[trim=0cm 0cm 0cm 0.8cm, clip, width=\textwidth]{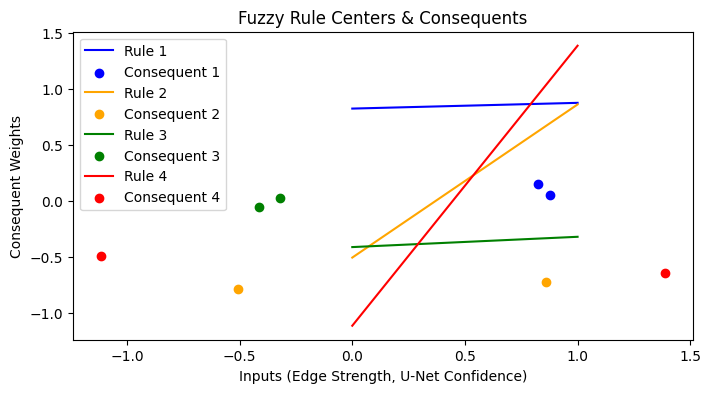}
        \caption{Rule-Base Inputs and Consequents}
        \label{fig:inputs_and_consequents}
    \end{subfigure}

    % --- Main Figure Caption and Label ---
    \caption{TSK Rule-Base Visualizations}
    \label{fig:rule_base_viz}
\end{figure}

\section{Discussion}

The defining feature of explainable sMoE U-Net is its ability to explain its predictions. We analyze this through two visualization mechanisms derived directly from the trained model.

\subsection{Logic Strategy Maps}
The "Strategy Maps" in Figure \ref{fig:strategy_maps} visualize the output of the gating network $G(E)$.
\begin{itemize}
    \item Observation: In homogeneous regions (e.g., the sky in Fig 5), the gate values approach 0, activating the Context Expert. This expert uses dilated convolutions to integrate broad spatial information, effectively suppressing noise.
    \item Observation: At object boundaries (e.g., the bear's outline), the gate values approach 1, activating the Boundary Expert. This expert uses pointwise convolutions to preserve sharp pixel transitions.
    \item Significance: This proves the network is actively switching strategies based on local image topology, rather than applying a "one-size-fits-all" convolution.
\end{itemize}
\subsection{Fuzzy Rule Firing Maps}
The "Rule Firing Maps" reveal which logical premises are active. By inspecting the learned centers and consequents in the figure, we can semantically label the rules:
\begin{itemize}
    \item Rule 1 (Strong Edge): Centers at high $x_1$ and $x_2$. This rule fires when both Sobel gradient and U-Net confidence are high. It represents "Unambiguous Edges."
    \item Rule 2 (Semantic Edge): Center at low $x_1$ but high $x_2$. This rule fires when the gradient is weak, but the U-Net is confident. This captures "Soft Boundaries" (e.g., visual contours with low contrast).
    \item Rule 3 (Noise Suppression): Centers at low values. This rule detects background and actively suppresses it.
    \item Rule 4 (Texture Suppression) : Centers at high $x_1$ (Edge Strength) but low $x_2$ (Semantic Confidence). The consequent weight for this rule is negative (as seen in Figure \ref{fig:inputs_and_consequents}), effectively suppressing false positive gradients to clean up the final output. For a deeper quantitative analysis of how these rules are formed, refer to the trained rules  visualization in Figure \ref{fig:x1_visual} and \ref{fig:x2_visual}.
\end{itemize}
This level of transparency allows domain experts to verify why an edge was detected. For example, if an edge is detected via Rule 2, the user knows it relies on learned semantic context rather than raw pixel intensity.

\section{Conclusion}
\label{sec:5}

The opacity of modern Deep Learning models remains a significant barrier to their adoption in verification-heavy fields. In this work, we presented the explainable sMoE U-Net, a hybrid architecture that bridges the gap between high-performance CNNs and interpretable fuzzy logic. By replacing standard convolutional layers with Spatially-Adaptive Mixture-of-Experts blocks, we enabled the network to dynamically switch between context-gathering and boundary-sharpening strategies. Furthermore, our novel TSK Fuzzy Head integration provides a transparent decision-making layer, allowing users to inspect the exact "IF-THEN" logic behind every pixel prediction.

Evaluation on the BSDS500 benchmark demonstrates no significant loss of accuracy due to transparency. The sMoE U-Net achieves an ODS F-score of 0.7628, outperforming the standard U-Net (0.7437) and matching specialized deep baselines like HED. Ultimately, this framework offers a viable path toward "Glass-Box" computer vision, where edge detection is both accurate and auditable.

\section{Future Work}
\label{sec:6}
While the current results are promising, several avenues for expansion remain. First, we aim to validate the sMoE U-Net in safety-critical domains, specifically medical imaging (e.g., tumor boundary definition) and aerospace maintenance (e.g., crack detection in jet engine blades), where the explainability of the TSK head offers a distinct advantage for regulatory compliance. Second, we plan to conduct rigorous ablation studies to quantify the individual contributions of the "Context" versus "Boundary" experts, ensuring the gating network is optimally utilized. Finally, we will explore extending the fuzzy logic head to higher-order TSK systems to capture even more complex edge dependencies without sacrificing interpretability.

\begin{credits}

\subsubsection{\discintname}
The authors have no competing interests to declare that are
relevant to the content of this article. 
\end{credits}

\begin{thebibliography}{8}

\bibitem{teng2024literature}Teng, Z., Li, L., Xin, Z., Xiang, D., Huang, J., Zhou, H., Shi, F., Zhu, W., Cai, J., Peng, T. \& Others A literature review of artificial intelligence (AI) for medical image segmentation: from AI and explainable AI to trustworthy AI. {\em Quantitative Imaging in Medicine and Surgery}. \textbf{14}, 9620 (2024)
\bibitem{lynn2021implementation}Lynn, N., Sourav, A. \& Santoso, A. Implementation of real-time edge detection using Canny and Sobel algorithms. {\em IOP Conference Series: Materials Science and Engineering}. \textbf{1096}, 012079 (2021)
\bibitem{cosgrove2020adversarial}Cosgrove, C. \& Yuille, A. Adversarial examples for edge detection: They exist, and they transfer. {\em Proceedings Of The IEEE/CVF Winter Conference on Applications of Computer Vision}. pp. 1070-1079 (2020)
\bibitem{dryden2022spatial}Dryden, N. \& Hoefler, T. Spatial mixture-of-experts. {\em Advances in Neural Information Processing Systems}. \textbf{35} pp. 11697-11713 (2022)
\bibitem{cai2025survey}Cai, W., Jiang, J., Wang, F., Tang, J., Kim, S. \& Huang, J. A survey on mixture of experts in large language models. {\em IEEE Transactions on Knowledge and Data Engineering}. (2025)
\bibitem{riquelme2021scaling}Riquelme, C., Puigcerver, J., Mustafa, B., Neumann, M., Jenatton, R., Susano Pinto, A., Keysers, D. \& Houlsby, N. Scaling vision with sparse mixture of experts. {\em Advances in Neural Information Processing Systems}. \textbf{34} pp. 8583-8595 (2021)
\bibitem{su_pixel_nodate}Su, Z., Liu, W., Yu, Z., Hu, D., Liao, Q., Tian, Q., Pietikäinen, M. \& Liu, L. Pixel difference networks for efficient edge detection. {\em Proceedings Of The IEEE/CVF International Conference on Computer Vision}. pp. 5117-5127 (2021)
\bibitem{xie2015holistically}Xie, S. \& Tu, Z. Holistically-nested edge detection. {\em Proceedings Of The IEEE International Conference on Computer Vision}. pp. 1395-1403 (2015)
\bibitem{rahim_towards_2025}Rahim, M. \& Mahmud, M. Towards Explainable Segmentation of Complex Boundaries in Lung Nodule Detection.  (2025,11), \url{https://openreview.net/forum?id=lOKyFN5wWm}
\bibitem{boskovitz_adaptive_2002}Boskovitz, V. \& Guterman, H. An adaptive neuro-fuzzy system for automatic image segmentation and edge detection. {\em IEEE Transactions on Fuzzy Systems}. \textbf{10}, 247-262 (2002,4), \url{https://ieeexplore.ieee.org/document/995125/}
\bibitem{ocak_iocak28unet_edge_detection_2024}Ocak, I. iocak28/UNet\_edge\_detection.  (2024,11), \url{https://github.com/iocak28/UNet_edge_detection}, original-date: 2020-06-04
\bibitem{ronneberger_u_net_2015}Ronneberger, O., Fischer, P. \& Brox, T. U-Net: Convolutional Networks for Biomedical Image Segmentation. (arXiv,2015,5), \url{http://arxiv.org/abs/1505.04597}, arXiv:1505.04597 [cs]
\bibitem{MartinFTM01}Martin, D., Fowlkes, C., Tal, D. \& Malik, J. A Database of Human Segmented Natural Images and its Application to Evaluating Segmentation Algorithms and Measuring Ecological Statistics. {\em Proc. 8th Int'l Conf. Computer Vision}. \textbf{2} pp. 416-423 (2001,7)
\end{thebibliography}
%

\end{document}